\documentclass[12pt]{article}
\usepackage{times}
\usepackage{xcolor}
\usepackage[version=4]{mhchem}
\usepackage{amsmath,stmaryrd,graphicx,amssymb}
\usepackage{mathtools}  
\usepackage{xfrac}  
\usepackage{nameref}
\usepackage{miller}
\usepackage{xr}
\usepackage[hidelinks]{hyperref}
\usepackage{gensymb}
\emergencystretch=\maxdimen
\usepackage[labelfont=bf]{caption}

\makeatletter
\renewcommand{\@seccntformat}[1]{%
  \ifcsname prefix@#1\endcsname
    \csname prefix@#1\endcsname
  \else
    \csname the#1\endcsname\quad
  \fi}

\makeatother

\usepackage[a4paper,
            left=0.75in,
            right=0.75in,
            top=0.75in,
            bottom=1in,
            footskip=.25in]{geometry}
            
\title{Insights into Chemical and Structural Order at Planar Defects in a Functional Oxide Using Multislice Electron Ptychography}

\author{Menglin Zhu$^{1\ast}$, Michael Xu$^{1\ast}$, Yun Yu$^{2}$, Liyan Wu$^{2}$, Or Shafir$^{3}$,\\Colin Gilgenbach$^{1}$, Lane W. Martin$^{4}$, Ilya Grinberg$^{3}$, Jonathan E. Spanier$^{2}$\\ James M. LeBeau$^{1\ast\ast}$\\\\
\normalsize{$^{1}$Department of Materials Science and Engineering, Massachusetts Institute of Technology,}\\
\normalsize{Cambridge, MA 02139, USA}\\
\normalsize{$^{2}$Department of Mechanical Engineering and Mechanics, Drexel University,}\\
\normalsize{Philadelphia, PA 19104, USA}\\
\normalsize{$^{3}$Department of Chemistry, Bar-Ilan University, Ramat Gan, 5290002, Israel}\\
\normalsize{$^{4}$Departments of Materials Science and NanoEngineering, Chemistry, and Physics and Astronomy}\\
\normalsize{and the Rice Advanced Materials Institute, Rice University, Houston, TX 77005, USA}\\\\
\normalsize{$^\ast$These authors contributed equally to this work.}\\
\normalsize{$^{\ast\ast}$To whom correspondence should be addressed; E-mail: lebeau@mit.edu.}}

\date{}

\begin{document}

\maketitle

\begin{abstract}
    
Switchable order parameters in ferroic materials are essential for functional electronic devices, yet disruptions of the ordering can take the form of planar boundaries or defects that exhibit distinct properties. Characterizing the structure of these boundaries is challenging due to their confined size and three-dimensional nature. Here, a chemical anti-phase boundary in the highly ordered double perovskite \ce{Pb2MgWO6} is investigated using multislice electron ptychography. The boundary is revealed to be inclined along the electron beam direction with a finite width of chemical intermixing. Additionally, regions at and near the boundary exhibit antiferroelectric-like displacements, contrasting with the predominantly paraelectric matrix. Spatial statistics and density functional theory calculations further indicate that despite their higher energy, chemical anti-phase boundaries form due to kinetic constraints during growth, with extended antiferroelectric-like distortions induced by the chemically frustrated environment in the proximity of the boundary. The three-dimensional imaging provides critical insights into the interplay between local chemistry and the polar environment, elucidating the role of anti-phase boundaries and their associated confined structural distortions and offering new opportunities for engineering ferroic thin films.

\end{abstract}

\newpage

\section{Introduction}

Ferroic materials, which exhibit spontaneous ordering of electric, magnetic, or elastic dipoles, are crucial for memory devices, sensors, and actuators due to their ability to switch states under external stimuli. Variations in the orientations of these order parameters lead to the formation of domains, separated by phase boundaries \cite{Tagantsev2010-yb}. When these boundaries involve a half-unit-cell shift between neighboring domains, they are known as anti-phase boundaries (APBs). Changes in order parameters at APBs can induce emergent properties, such as altered electric/magnetic response \cite{Asaka2007-xj, Wei2014-ir, McKenna2014-ua,Kumar2021-gu, Xu2022-rp}, domain switching \cite{Hensling2019-zf, Zhang2021-en}, and variations in electronic and ionic conductivity \cite{Nair2018-jr, Xu2023-ne}. Understanding and controlling the structure of these boundaries is therefore essential for optimizing the performance of ferroic materials and enhancing the functionality of related devices \cite{Wang2018-ua}.

The reduced dimensionality and localized changes associated with APBs require high spatial resolution for characterization. While scanning transmission electron microscopy (STEM) has been instrumental in revealing the local chemical and/or structural variations at planar boundaries \cite{Van-Aert2012-it, Farokhipoor2014-ft, Esser2016-gu, Xu2022-rp, Kumar2022-cm},  projection of such inherently three-dimensional defects onto two-dimensional images by the technique presents challenges in structural interpretation. For instance, this can obscure critical information of APBs, such as their inclination relative to the electron beam or their incommensurate nature. Multislice electron ptychography offers a solution to this issue by allowing depth-resolved information to be recovered without the need for through-focal series acquisitions or tilt tomography \cite{Chen2021-rc}. In this technique, electron diffraction patterns are recorded as the probe scans across the sample and used to iteratively reconstruct the sample potential slice by slice, offering additional depth resolution with atomic number sensitivity \cite{Jiang2018-as, Chen2021-rc, Dong2024-vk}. 

The compound \ce{Pb2MgWO6} (PMW) offers a model system for studying the three-dimensional structure of chemical-APBs (c-APBs)  using multislice electron ptychography. PMW is a double perovskite with general formula of \ce{A_2BB}'\ce{X_6}, in which lead and oxygen occupy the \ce{A} and \ce{X} sub-lattices, respectively, and antiparallel off-centered displacements of lead relative to its neighboring oxygens give rise to an antiferroelectric (AFE) ground state at room temperature ($T_c$ = 313 K). The B sub-lattice is divided into separate \ce{B} and \ce{B}' sites, alternately occupied by \ce{Mg^{2+}} and \ce{W^{6+}}, with a high degree of chemical order driven by the large disparity in their respective ionic size and valency \cite{Baldinozzi1993-wg, Baldinozzi1995-ex, Seshadri1999-hg, Samanta2022-re}. However, c-APBs, separating chemically ordered domains offset by one pseudocubic unit cell, may form during nucleation and growth depending on the substrate and process factors \cite{Vasala2015-fl, Spurgeon2016-ct}. These local deviations from the otherwise highly ordered structure allows for the study of c-APBs in isolation from other intrinsic defects, such as chemical fluctuations often observed in non-stoichiometric double perovskites \cite{Randall1990-if, King2010-vj}. This also facilitates investigating the impact of chemical environment on local structure and material functionality.

In this article, using a combination of multislice electron ptychography experiments and simulations of model structures, c-APBs in PMW are revealed to be inclined along an incommensurate plane with a finite width of chemical intermixing on the \ce{B} sub-lattice. At these chemically disordered boundaries, enhanced lead vs. oxygen displacements are observed, consistent with AFE-like polar order, in contrast to the overall paraelectric (PE) film. Additionally, c-APBs are found to couple with domain walls, disrupting the long-range order of antiferroelectric domains during PE-to-AFE phase transformation below $T_c$. Density functional theory (DFT) and Monte Carlo simulated annealing calculations further confirm that the enhanced lead polar displacements correlate with the heterogeneous magnesium/tungsten environment at the c-APBs. These findings demonstrate the capability of multislice electron ptychography to reveal the 3D nature of planar defects and offer insights into the formation of c-APBs and their potential for engineering (anti)ferroelectric structure and response.

\section{Results and Discussion}
The PMW thin films were epitaxially grown on \hkl(110)$_{O}$-oriented \ce{NdScO_3} (NSO) substrates using pulsed-laser deposition (details in \nameref{methods}), where the $O$ subscript specifies orthorhombic indexing. The resulting films exhibit high-quality epitaxy that are nearly matched to the pseudocubic lattice parameters of the NSO substrate (PE-PMW $a_{PC}=b_{PC}=c_{PC}=$400.3 pm, NSO $a_{PC}=c_{PC}=$402.8pm, $b_{PC}=$401.5pm; where the $PC$ subscript specifies pseudocubic indexing). 

\begin{figure}[!h]
\centering
  \includegraphics[width=6.5in]{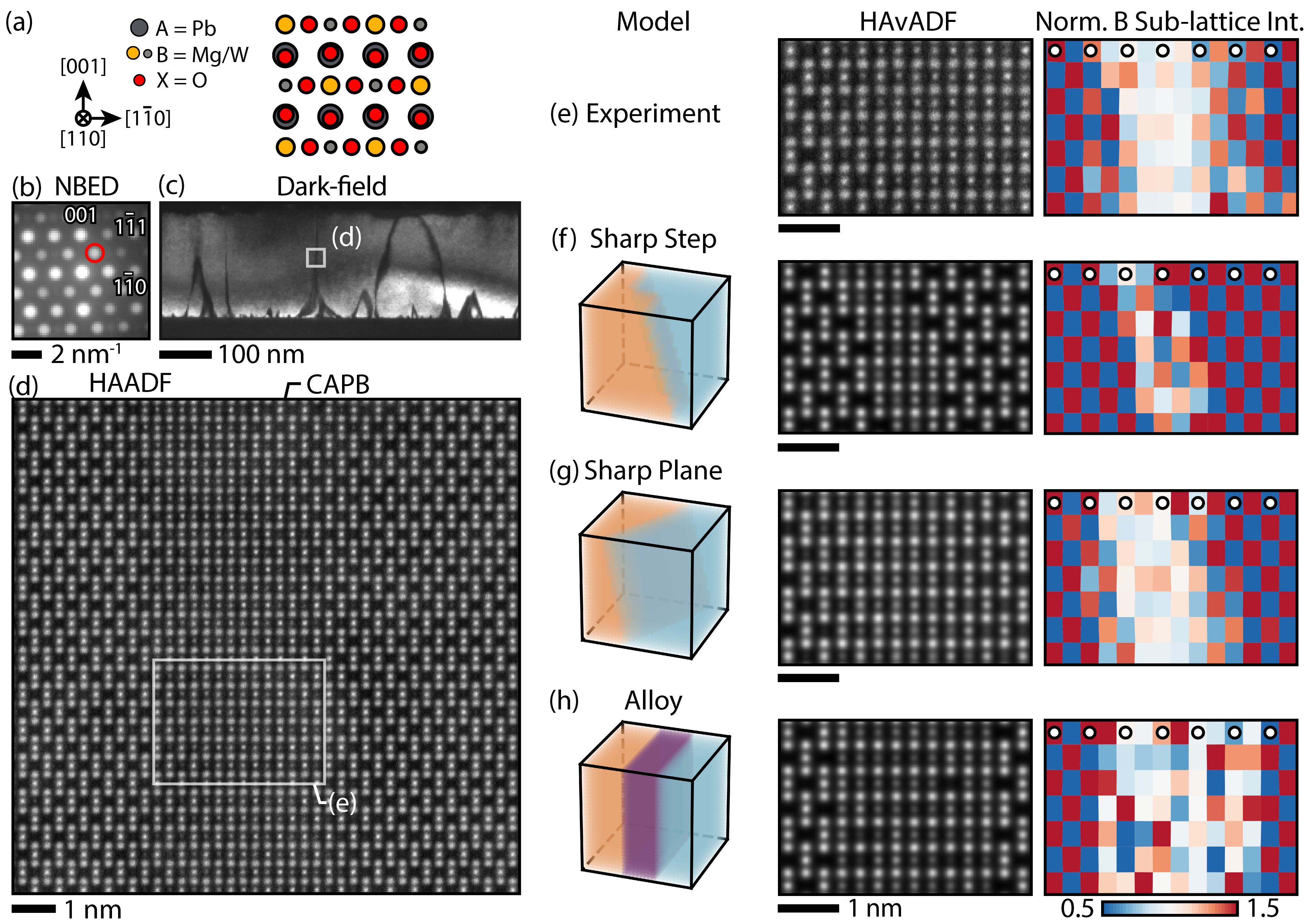}
  \caption{(a) Model of PE-PMW viewed along \hkl[110]$_{PC}$. (b) Average nano-beam electron diffraction pattern with the $\sfrac{1}{2}\,1\overline{1}1$ superlattice reflection marked. (c) Virtual dark-field image from the 4D STEM dataset formed using the $\sfrac{1}{2}\,1\overline{1}1$ superlattice reflections. (d) Corresponding HAADF image taken from one of the stripes with weak superlattice reflections in (c), showing a c-APB in the middle with different \ce{B} sub-lattice intensity compared to the surrounding domain, as confirmed by (e) the quantified normalized \ce{B} sub-lattice intensity. Schematic models of the c-APB separating domains with opposite chemical order vectors (red vs. blue), including (f) an atomically sharp boundary with a step along the depth, (g) a sharp boundary with an inclined orientation, and (h) a boundary with finite chemical intermixing (purple). Similar HAADF images simulated with these models and quantified \ce{B} sub-lattice intensities illustrate the ambiguity in interpreting the experimental image contrast.}
  \label{fig:haadf_comparison}
\end{figure}

When viewed along \hkl[110]$_{PC}$ (Figure~\ref{fig:haadf_comparison}a), PMW exhibits distinct atom columns containing lead/oxygen, oxygen, magnesium, or tungsten. This arrangement highlights the rocksalt ordering of magnesium/tungsten at the \ce{B} sub-lattice, which gives rise to the $\sfrac{1}{2}\,111$-type superlattice reflections circled in the nano-beam electron diffraction (NBED) pattern (Figure~\ref{fig:haadf_comparison}b) \cite{Baldinozzi1993-wg, Baldinozzi1995-ex, Anderson1993-wq, Yang1994-sa}. While the dark-field image (Figure~\ref{fig:haadf_comparison}c) formed with the $\sfrac{1}{2}\,1\overline{1}1$ reflection reveals an overall ordered film with uniform contrast, extended dark stripes and triangular-shaped features across the film suggest that the structure and/or chemistry of these extended defects deviate from the rest of the film structure. 
 The dark lines observed through the thickness of the film, however, persist throughout the temperature cycle, and the contrast of the AFE domains flips across some of these boundaries, resembling pinning effects in magnetically ordered double perovskites \cite{Asaka2007-xj, Meneghini2009-dl}. This suggests that although the presence of a chemical order change does not necessitate an AFE domain wall \cite{Randall1989-sc}, there may be a disrupting effect on the formation of long-range AFE order \cite{Yang1994-sa, Reaney2011-cy}.

Furthermore, although bulk PMW exhibits an AFE ground state at room temperature, its near-room-temperature Curie point, combined with substrate-induced strain and surface effects due to the finite size of a cross-section lamella, results in a predominantly PE phase at room temperature. During cryogenic cooling, additional features emerge in the dark-field image as AFE domains form (SI-Figure S1). The dark lines observed through the thickness of the film, however, persist throughout the temperature cycle, and the contrast of the AFE domains flips across some of these boundaries, resembling pinning effects in magnetically ordered double perovskites \cite{Asaka2007-xj, Meneghini2009-dl}. This suggests that although the presence of a chemical order change does not necessitate an AFE domain wall \cite{Randall1989-sc}, there may be a disrupting effect on the formation of long-range AFE order \cite{Yang1994-sa, Reaney2011-cy}.

\begin{figure}[!h]
\centering
  \includegraphics[width=3in]{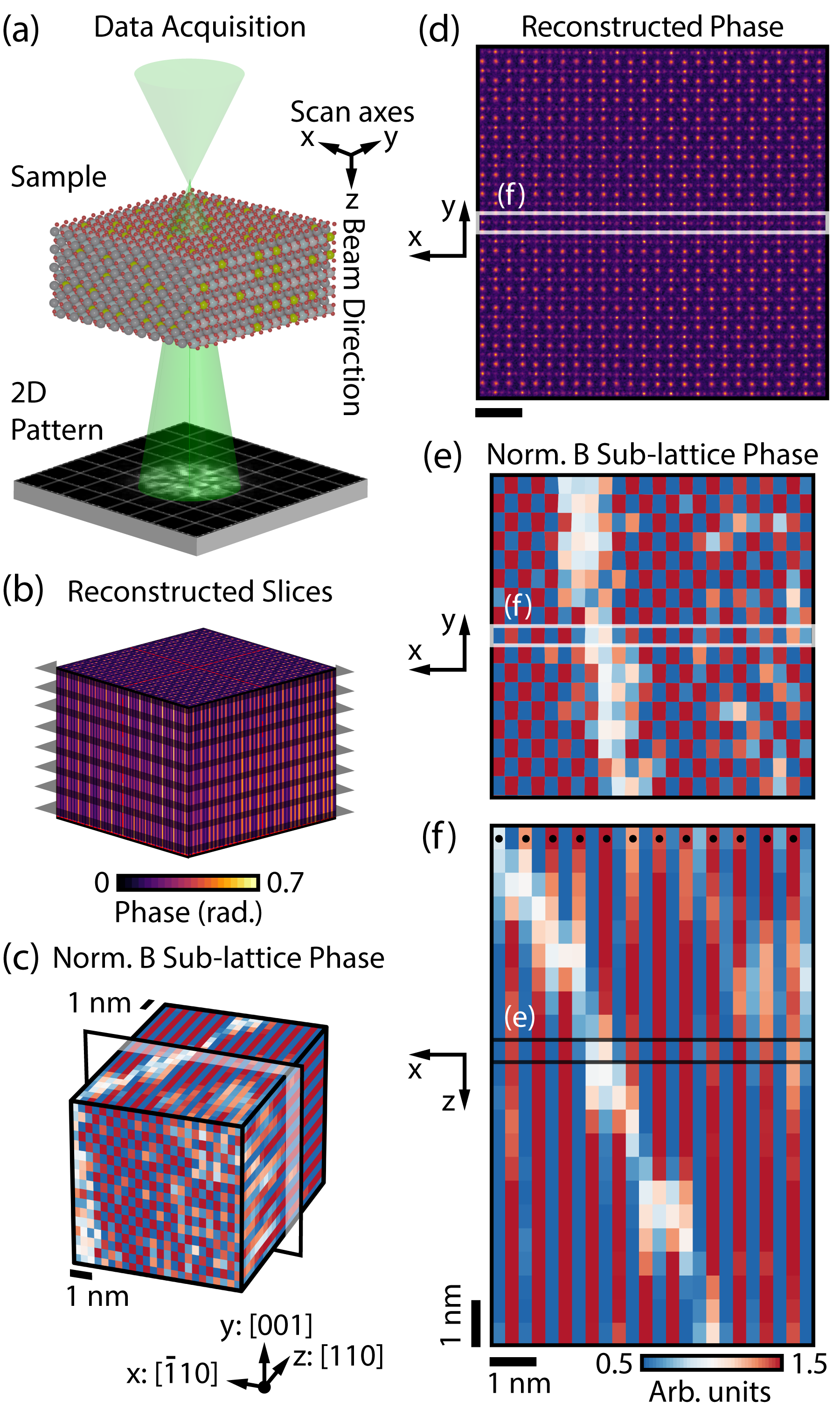}
  \caption{(a) Schematic overview of the ptychography experiment and (b) slice-wise reconstruction of the object phase along the \hkl[110] zone axis. (c) Volumetric normalized \ce{B} sub-lattice phase information extracted from each slice of the reconstructed object, with (d) an example phase image and (e) the corresponding \ce{B} sub-lattice phase shown. The position of the slice is indicated by the box in (c). (f) Cross-section along the view direction, extracted from the boxed region in (e), reveals the inclination of the c-APB along the depth.}
  \label{fig:ptycho_result}
\end{figure}

Closer examination of the stripes (boxed region in Figure~\ref{fig:haadf_comparison}c) using atomic number-sensitive HAADF STEM imaging (Figure~\ref{fig:haadf_comparison}d) reveals alternating bright and near-absent atom columns on the \ce{B} sub-lattice, reflecting rocksalt-ordered tungsten ($Z=74$) and magnesium ($Z=12$). This ordering, however, exhibits a half-unit-cell shift across the boundary in the middle of the image, indicating the presence of a c-APB. Furthermore, the HAADF intensities of \ce{B} sub-lattice atom columns at the c-APB are uniformly distributed, differing from those of either side of the domains. Considering the projective averaging of HAADF imaging, this anomalous contrast may be a result of either chemical intermixing between magnesium and tungsten or the morphology of the boundary itself \cite{Esser2016-gu}. This point is illustrated by comparing experiment (Figure~\ref{fig:ptycho_result}e) with STEM simulations of three distinct boundary models (Figure~\ref{fig:haadf_comparison}f-h), which yield HAADF image and normalized \ce{B} sub-lattice intensities that are all visually similar (details in Experimental Section). For the same reason, identifying any structural and property changes driven by the c-APB remains challenging due to projection of the sample \cite{Calderon2022-ss}. 

To overcome the inherent ambiguities of HAADF STEM imaging, multislice electron ptychography is used. This technique uses the interference between overlapping convergent beam electron diffraction patterns acquired across the sample (Figure~\ref{fig:ptycho_result}a) to iteratively reconstruct the depth-resolved sample phase (Figure~\ref{fig:ptycho_result}b), while accounting for dynamical scattering through a multislice approach \cite{Chen2021-rc}. As a result, the structural and chemical environment of the c-APB can be elucidated in 3D. 

A representative slice of the reconstructed object (Figure~\ref{fig:ptycho_result}d) obtained from multislice electron ptychography (details in Experimental Section) shows significantly enhanced atomic-number sensitivity and spatial resolution compared to the HAADF image of the same region (Figure~\ref{fig:haadf_comparison}d). For example, both lead/oxygen and oxygen atom columns are resolved with an information limit below 40 pm (SI-Figure S2b) \cite{Chen2021-rc}. Mapping of the normalized \ce{B} sub-lattice object phase (Figure~\ref{fig:ptycho_result}e) from this slice reveals rocksalt ordering of magnesium and tungsten on either side of the image, with an half-unit-cell shift between them. Additionally, the \ce{B} sub-lattice atom columns in the c-APB region exhibit uniform phase within a finite width, consistent with observations from HAADF imaging yet significantly narrower in extent. Quantification of the \ce{B} sub-lattice object phase across the entire ptychographic image stack further reveals the 3D structure of the boundary (Figure~\ref{fig:ptycho_result}c, f), which is shown to be inclined with respect to the view direction (\hkl[110]$_{pc}$). Likewise, the B sub-lattice atom columns on each 5-Å-thick reconstructed object slice show a uniform phase within the 2–5 unit cell ($\sim$3–11 Å) boundary width. 

\begin{figure}[!h]
\centering
  \includegraphics[width=3in]{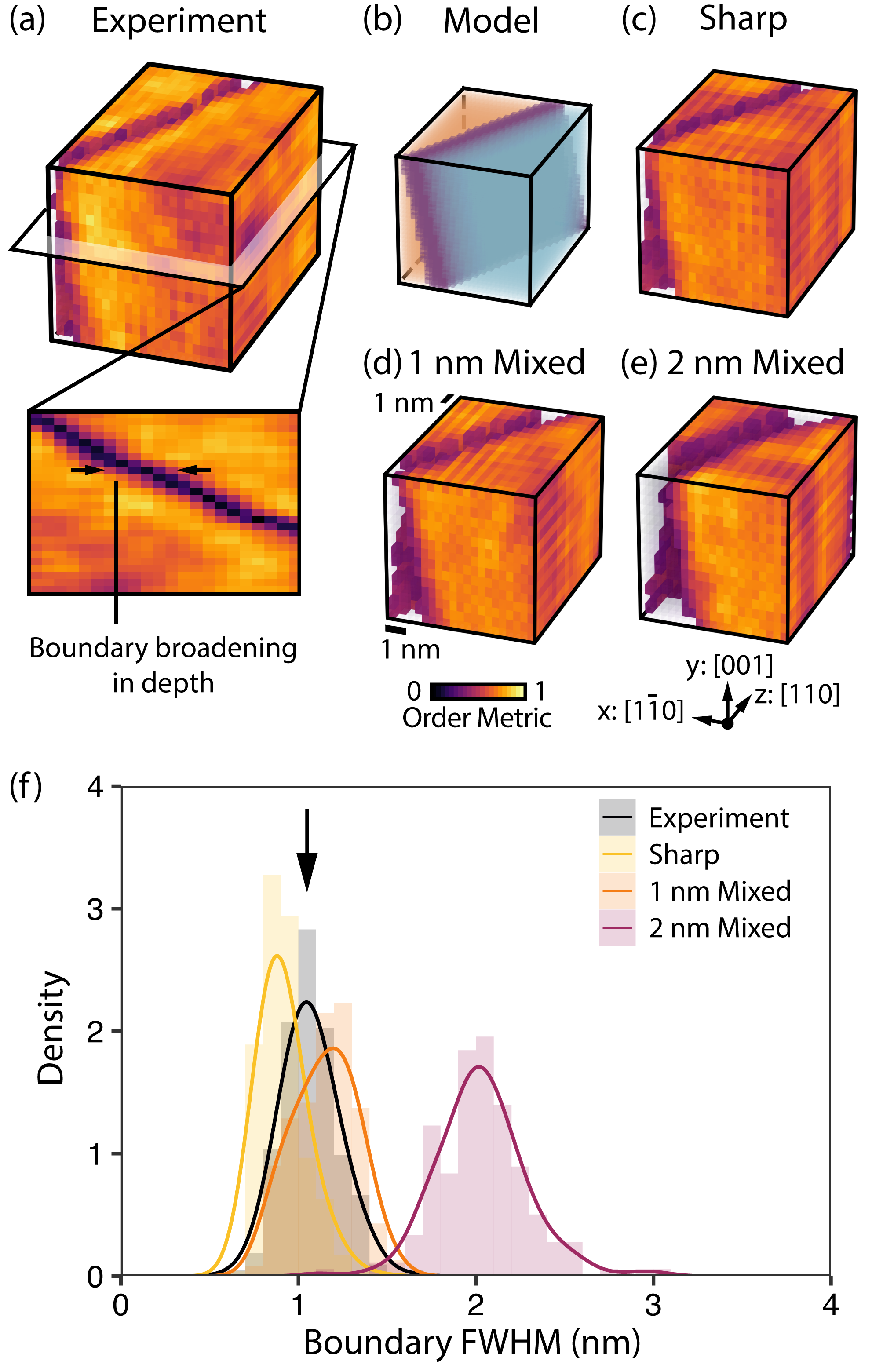}
  \caption{(a) Cross-sectional and 3D views of the experimental chemical order metric. (b) A plane is fitted to the region with a low chemical order metric, and crystal models are constructed by assigning intermixed regions of varying widths along the plane (purple) to separate domains with opposite chemical order vectors (red vs. blue). The 3D views of the chemical order metric are presented for models with (c) atomically sharp, (d) 1 nm, and (e) 2 nm intermixed boundaries. (f) Distribution of the \textit{apparent} width of c-APB, determined from the FWHM of a fitted pseudo-Voigt function for the order metric of the atomic column along the view direction ($z$). }
  \label{fig:chem_order_cubes}
\end{figure}

The extent of \textit{apparent} chemical disorder in the heterogeneous region is further quantified by correlation analysis in 3D (methods in SI) \cite{Kumar2021-gu}. In areas of chemical disorder (c-APB), the correlation result is lower (Figure~\ref{fig:chem_order_cubes}a, represented by transparency) than in the chemically ordered thin film ``bulk.'' On the other hand, although a lower order metric, or equivalently uniform \ce{B} sub-lattice object phase, might suggest a boundary with chemical intermixing \cite{Egoavil2015-ct}, the limited depth resolution of multislice electron ptychography introduces additional ambiguity in determining the depth-wise occupancy of magnesium and tungsten across the c-APB. To elucidate the effects of limited depth resolution and the origin of the near-uniform \ce{B} sub-lattice object phase observed in experiment, comparisons are made with reconstruction of model structures. The orientation of the c-APB is first established by fitting a plane to the region with intermediate object phase. The plane of the boundary is found to be inclined at $\sim$64$\degree$ relative to the view direction with slight curvature, and is approximately on \hkl(3 1 1). Using this reference plane, ptychography datasets are subsequently simulated and reconstructed for three model structures: one with an atomically sharp boundary between the domains and two others with 1 nm and 2 nm intermixed regions (Figure~\ref{fig:chem_order_cubes}b). Finally, to measure the depth extent of the c-APB, a pseudo-Voigt function is fit to the order metrics of each atomic column along the viewing direction for the experiment and simulated reconstruction stacks (Figure~\ref{fig:chem_order_cubes}a, c-e). The \textit{apparent} boundary width is represented by the full width at half maximum (FWHM) (Figure~\ref{fig:chem_order_cubes}f).

Due to the limited depth resolution, the average FWHM for the atomically sharp boundary is 0.910$\pm$0.126 nm (0.455$\pm$0.063 5-Å slices), while boundaries with 1 and 2 nm widths have average FWHMs of 1.15$\pm$0.173 and 2.03$\pm$0.254 nm, respectively. Comparison with simulations suggests that the experimental data, with an average FWHM of 1.07$\pm$0.153 nm, represents an intermediate scenario. Some regions of the boundary are atomically sharp along the depth, while others exhibit a few unit cells of intermixing. This variability is also evident from the depth sectioning shown in Figure~\ref{fig:ptycho_result}f.

\begin{figure}[!h]
\centering
  \includegraphics[width=6.5in]{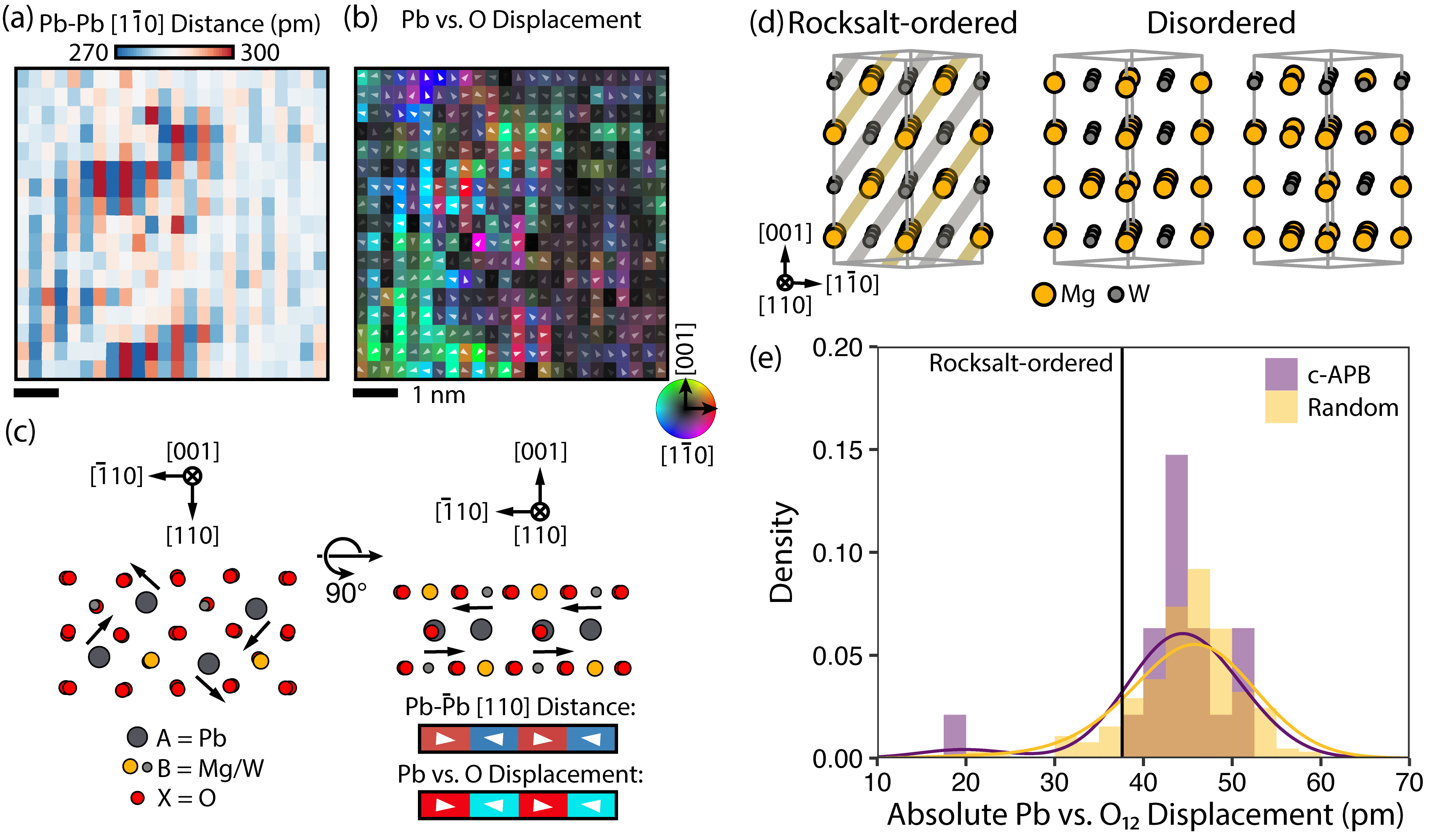}
  \caption{(a) \ce{Pb}-\ce{Pb} distance along \hkl[1-10] and (b) projected polar displacement measured between the mean position of lead and its four nearest neighboring oxygen atoms, as measured from the same slice in Figure~\ref{fig:ptycho_result}e. (c) Schematic model of the PMW unit cell viewed from the \hkl[001] and \hkl[110] (experimental) zone axes, with arrows indicating the antiparallel displacement of lead and the resulting alternating long-short \ce{Pb}-\ce{Pb} distances and polar displacement observed in the ptychography slice. (d) Schematic models of different supercells used for full DFT calculations, along with (e) the distribution of their respective absolute polar displacements measured between lead and its twelve neighboring oxygen atoms.}
  \label{fig:afe_displacements_dft}
\end{figure}

With the 3D structure of the c-APB determined, repeated unit cells of \ce{Pb2W2O6} and \ce{Pb2Mg2O6} along the atomically sharp boundary, as well as non-rock-salt ordered \ce{B} sub-lattice configurations in the intermixed region, are expected to generate chemical pressure and significantly alter the polar environment \cite{Ting2006-jo, Vasala2015-fl, Morrow2016-wj, Dey2017-mb, Burkert2017-cc}. Leveraging the high lateral resolution and depth-sensitive information offered by ptychography, structural distortions are quantified by mapping the lead-lead \hkl[1-10] distances (Figure~\ref{fig:afe_displacements_dft})a) for the same slice in Figure~\ref{fig:ptycho_result}d. While the field of view largely shows uniform lead-lead distances, an alternating pattern of long and short distances emerges near the c-APB, suggesting a localized change in bonding environment.

Comparison with the atomic model of AFE-PMW along one of the \hkl<110>$_{pc}$ (Figure~\ref{fig:afe_displacements_dft})c, and SI-Figure S4) reveals a similar pattern of lead-lead distances, attributed to the projected antiparallel displacements of neighboring lead/oxygen columns along the \hkl[-110]$_{pc}$ and \hkl[1-10]$_{pc}$ directions. With anions considered, the projected polar displacement between oxygen and lead, mapped from the same experimental slice, shows a comparable striped pattern near the c-APB (Figure~\ref{fig:afe_displacements_dft}b). Additional lead distortions of nearly 20\% (4.2 $\pm$ 2.5 pm vs. 3.5 $\pm$ 2.0 pm), represented by an increase in the average magnitude of projected polar displacements, are found at the c-APB as well. This suggests that both increased structural distortions and stronger AFE-like polar ordering persist near the c-APB, despite the majority of the thin film exhibiting a PE-like non-polar structure.

To further elucidate the energetics and polar ordering associated with c-APBs, density functional theory (DFT) calculations are employed to complement the experimental observations. The effect of magnesium/tungsten arrangement on energy is investigated by performing DFT simulations on 200 randomly generated 80-atom supercells ($2\times2\times4$ pseudocubic unit-cells), each featuring distinct \ce{B} sub-lattice cation arrangements. The extent of \ce{B} sub-lattice orderings is represented by 10 different four-atom chain configurations. To facilitate calculation on large supercells that better represent the c-APB formation, an energy regression model is constructed for the PMW supercell as a function of the counts of various \ce{B} sub-lattice chains (details in Experimental Section and SI). Leveraging the energy model, Monte Carlo simulated annealing with different starting temperatures and cooling rates is performed on large \ce{B} sub-lattice-only supercells ($16\times16\times16$ pseudocubic unit-cells) to explore c-APB formation and plausible low-energy magnesium/tungsten arrangements. In contrast to a fully rocksalt-ordered magnesium/tungsten supercell, which would form under infinitely slow cooling, several boundaries with low-energy deviations from rocksalt order emerge under realistic cooling rates. These boundaries, separating regions with opposite rocksalt order vectors, represent local disruptions of order parameters observed at the c-APB in the experiment. 

The energy and structure of these c-APBs are investigated through full DFT relaxations on 80-atom snapshots extracted from the Monte Carlo-annealed supercells (Figure~\ref{fig:afe_displacements_dft}d). The results are compared to those from a similar supercell featuring randomly generated \ce{B} sub-lattice occupancy without annealing. Since \ce{B} sub-lattice arrangements deviating from rocksalt order are less stable and have higher relative energy, the annealed supercell shows a greater prevalence of rocksalt ordering near the boundary. Consequently, the average relative energy of the boundary in the annealed supercell is lower compared to that in the random configurations ($34.2 \pm 15.0$ meV/atom vs. $52.9 \pm 19.9$ meV/atom) (SI-Figure S7). Structurally, the absolute off-center displacements of lead versus its 12 nearest neighbor oxygens are $\sim$15\% higher ($43 \pm 7$ pm) around the boundary in both the random and annealed supercells compared to those in the rocksalt-ordered PMW (37 pm) (Figure~\ref{fig:afe_displacements_dft}e). This increased displacement, consistent with the relative difference in projected displacements from experiment, indicates that the disrupted chemical ordering can promote stronger AFE-like ordering by inducing correlated structural distortions. 

Despite being higher in energy, chemically frustrated boundaries inevitably form due to nucleation and growth during film fabrication \cite{Wang2018-ua, Ning2021-mh}. The increase of polar displacement in presence of chemical frustration can be ascribed to more highly-over-bonded (in the case of \ce{W^{6+}-O-W^{6+}}) or under-bonded (\ce{Mg^{2+}-O-Mg^{2+}}) oxygens near the boundary \cite{Grinberg2004-ac}. These oxygens, in turn, exert stronger repulsion (over-bonded) or attraction (under-bonded) on lead. For the \ce{B} sub-lattice cations, however, their displacement differences are found to be negligible (SI-Figure S7), suggesting cooperative \ce{BO_6} distortions as seen in other complex oxides \cite{Sang2015-nj}. 

\begin{figure}[!h]
\centering
  \includegraphics[width=3in]{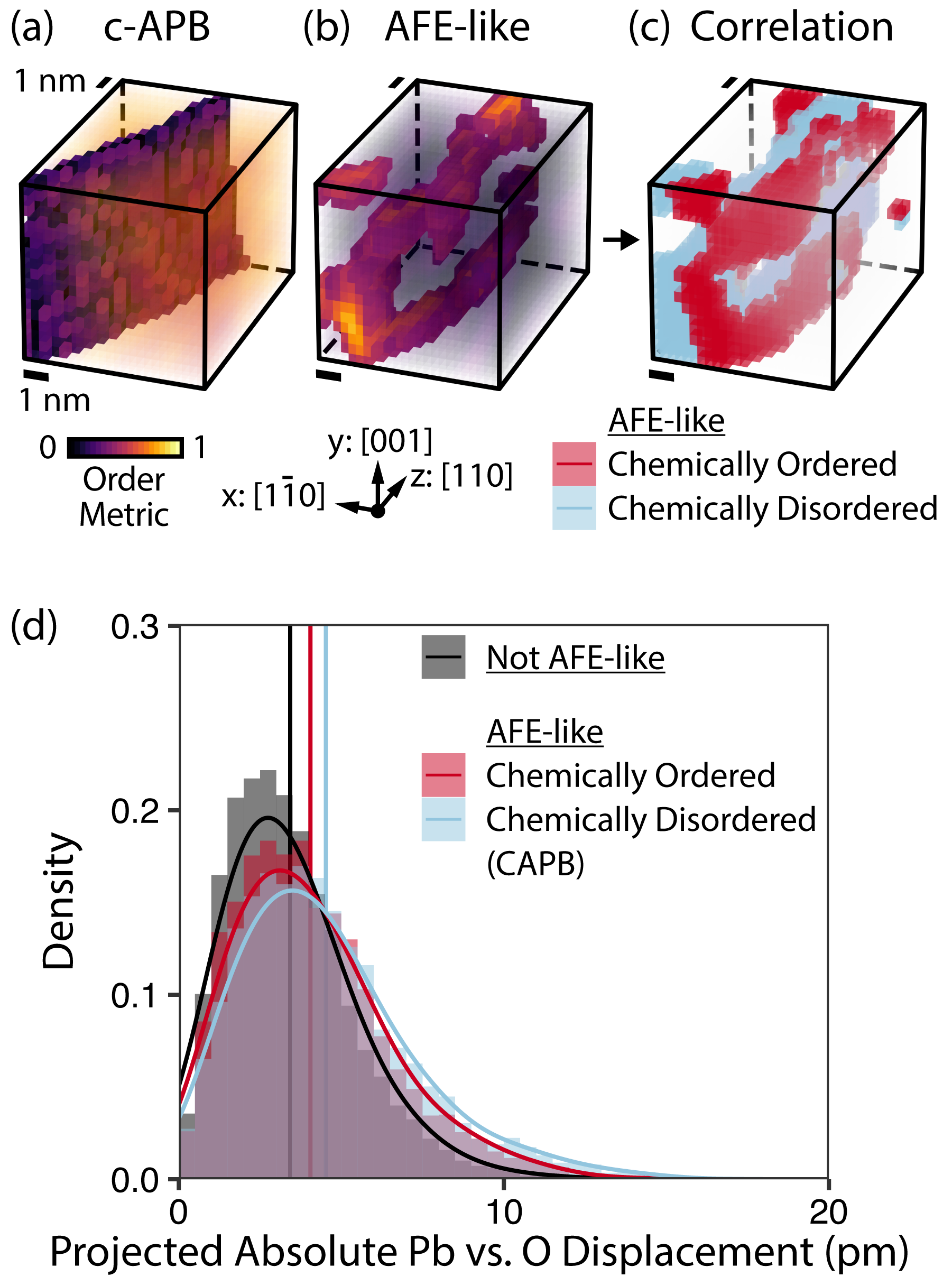}
  \caption{Volumetric view of (a) the chemical order metric, with solid color indicating the inclined and chemically disordered c-APB, and (b) the AFE-like ordering metric, with solid color highlighting nanodomains characterized by alternating long-short \ce{Pb}-\ce{Pb} distances. (c) The nanodomains are clustered into those situated directly on the chemically disordered c-APB (blue) and those near the c-APB but within the chemically ordered domain (red). (d) The distribution of their respective projected polar displacements.}
  \label{fig:correlation_chem_afe}
\end{figure}

The structural analysis of polar ordering can be further extended in 3D using the ptychography reconstruction stack. To quantify the alternating long-short lead-lead distances that are indicative of AFE ordering, a slice-wise correlation analysis is conducted using a striped kernel to yield an AFE-like order metric. This analysis reveals nanometer-sized structurally ordered domains extending in 3D (Figure~\ref{fig:correlation_chem_afe}b). Although these AFE-like nanodomains are not perfectly aligned with the c-APB, they are situated in close proximity to the boundary. By categorizing them into regions directly at or near the c-APB (blue and red in Figure~\ref{fig:correlation_chem_afe}c, SI-Figure S6, and further details in SI), it is evident that the polar displacements in these regions are much larger than those in the matrix, with the chemically disordered regions exhibiting the highest displacements (Figure~\ref{fig:correlation_chem_afe}d). This observation highlights the critical role that chemical order/disorder has in enhancing polar distortions and facilitating structural order, with a ripple effect extending beyond the c-APB. Further into the chemically ordered regions, however, the AFE-to-PE structural relaxation of the ``bulk'' PMW film prevents the formation of a long-range AFE phase. 

The formation of AFE-like nanoregions is akin to the polar nanodomains introduced by heterogeneities in lead-based relaxor ferroelectrics \cite{Cowley2011-is, Takenaka2017-sj, Kumar2021-gu}. However, in contrast to the diffuse order/disorder in relaxors, the chemical disorder at c-APBs is more limited in extent and dimensionality due to the strong preference for magnesium/tungsten ordering in PMW. Therefore, any structural changes arising from charge imbalance and chemical mismatch are proximate to the c-APB region, as corroborated by both experimental and theoretical results. Such a defined and localized effect enables another degree of control of polar structure in PMW thin films, including interactions with AFE domain walls observed at cryogenic conditions (SI-Figure S1). The distribution of c-APBs can be effectively manipulated through substrate surface steps, cation off-stoichiometry, oxygen vacancies, and lattice symmetry mismatch with the substrate \cite{Wang2018-ua}. This offers opportunities to precisely manipulate chemical order parameter and, consequently, local ferroic structure \cite{Wei2014-ir} and device behavior \cite{Zhang2021-en} by adjusting film composition, defect morphology, or defect density during growth. 

\section{Conclusion}

Leveraging the depth sensitivity of multislice electron ptychography and systematic comparison with image simulations, c-APBs in PMW thin films are shown to be characterized by an inclined orientation and a finite width of chemical intermixing. DFT calculations further elucidate that, despite their high-energy nature, c-APBs form due to kinetic-limited growth processes and induce significant structural changes that enhance polar distortions. This effect is confirmed experimentally, originating near the c-APBs and leading to the formation of antiferroelectric-like nanodomains, even as the bulk material relaxes to a paraelectric structure. Compared to heterogeneity-driven polar nanodomains observed in lead-based relaxor ferroelectrics, c-APB-induced changes in PMW are spatially confined and also directly affected by growth parameters, presenting a unique opportunity for tailoring local film structure through precise defect manipulation. These findings highlight the utility of multislice electron ptychography determining the 3D film structure near planar defects, in turn providing insight into how local disruptions in order can affect the structure and therefore properties of ferroic materials.

\section{Experimental Section}
\label{methods}

\subsection*{Thin Film Growth}
Epitaxial PMW thin films with a thickness of 200 nm were grown on \hkl(110)$_{O}$-oriented \ce{NdScO3} substrates using a pulsed layer deposition system (TSST). A KrF excimer laser (248 nm, Coherent) was used for deposition with a target-to-substrate distance of 5.5 cm. The lead-rich (10\% excess) PMW target was fabricated by solid-state reaction. The growth temperature and oxygen partial pressure were 580 $^\circ$C and 200 mTorr, respectively. A laser fluence of 2 J/cm$^2$ and a laser repetition rate of 2 Hz were utilized for film growth. After the deposition, the sample was cooled from growth temperature to room temperature at 5 $^\circ$C/min under a static oxygen pressure of ~400 Torr.

\subsection*{Scanning Transmission Electron Microscopy}
Cross-section samples for STEM characterization were prepared by \ce{Ga+} focused ion beam (Thermo Fisher Scientific Helios 600) following the conventional lift-out method. Thinning to electron transparency was achieved by milling with decreasing \ce{Ga+} ion beam energy from 30 to 5 kV. Final thinning to remove \ce{Ga+} damage was performed by broad beam \ce{Ar+} ion milling (Fischione 1051 TEM Mill) using beam energies of 0.3 and 0.1 kV. Electron microscopy datasets were acquired using a probe aberration-corrected STEM (Thermo Fisher Scientific Themis-Z S/TEM) at an accelerating voltage of 300 kV with probe semi-convergence angle of 21.8 mrad and beam current of $\approx$10 pA.

\subsection*{Multislice Electron Ptychography}
Datasets for ptychographic reconstructions were acquired using the Electron Microscope Pixel Array Detector (EMPAD) \cite{Tate2016-hd} with a dwell time of 1 ms per pixel and 10 nm overfocus. The 4D STEM dataset consists of 2D-$128\times128$ diffraction patterns collected over an array of $256\times256$ scan positions. 
Multislice electron ptychography reconstructions were performed using code based on PtychoShelves \cite{Chen2021-rc, Wakonig2020-aa}. The crystal lattice is divided into a stack of 46 slices, each with a thickness of 0.5 nm. Utilizing an iterative phase retrieval algorithm, the structure of the stack is reconstructed to match the measured diffraction intensities. 

\subsection*{STEM Image Simulations}
Simulation of 4D STEM and HAADF data were carried out using a Python-based implementation of the multislice method. The simulation parameters, including acceleration voltage and convergence angle, were selected to match the experimental conditions. Thermal vibrations of atoms were incorporated using the frozen phonon approximation. The finite effective source size was accounted for by randomly displacing the probe positions from the ideal grid for each frozen phonon configuration. The displacements were selected from a 60 pm full-width at half-maximum Gaussian distribution. Additional details can be found in SI.

\subsection*{Density Functional Theory Calculations}
The DFT calculations were performed using the Quantum ESPRESSO \cite{Giannozzi2009-zi} package, using the pseudopotentials from the GBRV \cite{Garrity2014-vp} pseudopotential database and the local density approximation (LDA) \cite{Parr1995-no, Perdew1981-wf} exchange-correlation functional. The ionic relaxations and electronic structure evaluation were performed on $2\times2\times1$ and $4\times4\times2$ Monkhorst-Pack k-point grids, respectively. For the parameterization of the energy model as a function of \ce{B} sub-lattice arrangement, we employed an 80\%/20\% training/test set split five times and subsequently combined the results of the obtained five models into a single final model for use in the Monte Carlo simulations. More details of the calculations can be found in SI.

\section*{Acknowledgements}
This research was sponsored by the Army Research Laboratory and was accomplished under Cooperative Agreement Number W911NF-24-2-0100. The views and conclusions contained in this document and those of the authors should not be interpreted as representing the official policies, either expressed or implied, of the Army Research Laboratory or the U.S. Government. The U.S. Government is authorized to reproduce and distribute reprints for Government purposes, notwithstanding any copyright notation herein. L.W.M. and J.E.S. also acknowledge the support of the Army Research Office under W911NF-21-1-0126. This work made use of the MIT.nano Characterization Facilities. 

\section*{Author Information} 
\subsection*{Contributions}
M.Z. and M.X. conducted electron microscopy experiments. M.Z., M.X., and C.G. conducted data analysis and image simulations. Y.Y. and L.W. grew the PMW thin films. O.S. performed DFT calculations. J.M.L., J.E.S., I.G., and L.W.M. designed and supervised this research. All authors co-wrote and edited the manuscript.

\section*{Ethics declarations}
\subsection*{Competing Interests}
The authors declare no competing interests.

\section*{Data Availability} 
The 4D STEM datasets, hyperparameters used for reconstruction, and the final output can be accessed at \url{Zenodo} upon acceptance of the manuscript. Other data and codes for image analysis are available from the corresponding author by reasonable request.

\bibliographystyle{MSP}
\bibliography{paperpile.bib}
\end{document}